\documentclass[aps,prl,twocolumn,showpacs,floatfix]{revtex4-1}

\usepackage{graphicx,amsmath}

\usepackage{amssymb}

\newcommand{\Ca}{CaFe$_2$As$_2$}
\newcommand{\Ba}{BaFe$_2$As$_2$}

\begin{document}

\title{Uniaxial versus hydrostatic pressure-induced phase transitions in CaFe$_2$As$_2$ and BaFe$_2$As$_2$}

\author{Milan Tomi{\'c}}

\affiliation{Institut f\"ur Theoretische Physik, Goethe-Universit\"at Frankfurt, Max-von-Laue-Stra{\ss}e 1, 60438 Frankfurt am Main, Germany}

\author{Roser Valent\'\i}

\affiliation{Institut f\"ur Theoretische Physik, Goethe-Universit\"at Frankfurt, Max-von-Laue-Stra{\ss}e 1, 60438 Frankfurt am Main, Germany}

\author{Harald O. Jeschke}

\affiliation{Institut f\"ur Theoretische Physik, Goethe-Universit\"at Frankfurt, Max-von-Laue-Stra{\ss}e 1, 60438 Frankfurt am Main, Germany}

\date{\today}

\begin{abstract}
  We present {\it uniaxial} pressure structural relaxations for {\Ca}
  and {\Ba} within density functional theory and compare them with
  calculations under hydrostatic pressure as well as available
  experimental results.  We find that {\Ca} shows a unique phase
  transition from a magnetic orthorhombic phase to a nonmagnetic
  collapsed tetragonal phase for both pressure conditions and no
  indication of a tetragonal phase is observed at intermediate
  uniaxial pressures.  In contrast, {\Ba} shows for both pressure
  conditions two phase transitions from a magnetic orthorhombic to a
  collapsed tetragonal phase through an intermediate nonmagnetic
  tetragonal phase.  We find that the critical transition pressures
  under uniaxial conditions are much lower than those under
  hydrostatic conditions manifesting the high sensitivity of the
  systems to uniaxial stress.  We discuss the origin of this
  sensitivity and its relation to experimental observations.
\end{abstract}









\maketitle

The discovery of superconductivity in iron
pnictides~\cite{Kamihara2008} has initiated an enourmous amount of
activities related to these materials.  Superconductivity can be
triggered either by chemical doping or by application of pressure
on the undoped parent compounds. One of the families that has been
intensively studied under pressure is the 122 family
$AE$Fe$_2$As$_2$ ($AE$ = Ca, Sr, and Ba).  {\Ca} at ambient pressure
undergoes a first order phase transition from a tetragonal to an
orthorhombic phase  at 172~K accompanied by a magnetic
transition. Initial reports on pressure experiments showed that at
$P\sim 0.23$~GPa the orthorhombic and antiferromagnetic phases are
suppressed and the system superconducts at low
temperatures~\cite{Torikachvili2008,Kreyssig2008}. Moreover, a
compressed tetragonal phase -- also called 'collapsed' tetragonal
phase --  was identified at higher pressures.  Subsequent
susceptibility and transport measurements under hydrostatic conditions
showed at low temperatures and $P \sim 0.35$~GPa a sharp orthorhombic
to collapsed tetragonal phase but no signature of
superconductivity~\cite{Yu2009}. In contrast, recent neutron
diffraction experiments on {\Ca} under uniaxial pressure along the $c$
axis~\cite{Prokes2010} indicate for pressures above 0.06~GPa and low
temperatures the presence of an intermediate nonmagnetic tetragonal
phase between the magnetic orthorhombic and the nonmagnetic collapsed
tetragonal phases. This phase was identified by the authors as the
phase responsible for superconductivity at $T=10$~K.
Other reports based on muon spin-relaxation measurements suggest
the existence of superconductivity in the orthorhombic phase,
raising the question whether superconductivity and magnetism
can coexist~\cite{Goko2009}.

{\Ba} shows an even more complex behavior under pressure.  At ambient
pressure it undergoes a phase transition from a metallic tetragonal
phase to an orthorhombic antiferromagnetic phase at $T=140$~K. Under
pressure the gradual appearance of a superconducting dome has been
observed by various groups~\cite{Alireza2009} though the role of
nonhydrostatic conditions is not yet well
understood~\cite{Paglione2010}.  Recent synchrotron X-ray diffraction
experiments under pressure~\cite{Uhoya2010} observe
at $T=300$~K a tetragonal to collapsed tetragonal phase
transition at $P= 22$~GPa under hydrostatic conditions
while this transition appears already at $P= 17$~GPa under
nonhydrostatic conditions.  On the other hand, 
 the authors of Ref.~\cite{Mittal2011} find at a lower temperature of
$T=33$~K  that 
{\Ba} undergoes a phase transition from a magnetic orthorhombic to a
nonmagnetic collapsed tetragonal phase at $P= 29$~GPa and report an
anomaly in the As-Fe-As bond angles at 10~GPa that they ascribe to be
of electronic origin. In contrast, high-pressure neutron diffraction
experiments~\cite{Kimber2009} performed at $T=17$~K find a tetragonal
phase at 3~GPa and 6~GPa.

These experimental results show that the onset of superconductivity as
well as the appearance of several structural phases at low
temperatures in {\Ca} and {\Ba} are extremely sensitive to the
pressure
conditions~\cite{Goldman2009,Prokes2010,Yamazaki2010,Duncan2010} and
are a subject of intensive debate.  In view of this strong controversy
we present {\it ab initio} density functional theory results for the
electronic, magnetic and structural behavior of both systems under
uniaxial and hydrostatic pressure conditions.  Previous theoretical
approaches which have examined the properties of the 122 family under
hydrostatic pressure have employed either fixed volume structural
optimizations~\cite{Colonna2011} or molecular
dynamics~\cite{Zhang2009}. Recently, anisotropic pressure studies
on {\Ba}
based on ground state
geometry calculations of more than 300 structures at different
fixed volumes were
reported in Ref.~\cite{Colonna2011_2}.
 Our approach consists of constant pressure
structural relaxations which allows us to probe the low-temperature
portion of the phase diagram in a relatively simple and
straightforward way.  With this approach we can treat {\it
  nonhydrostatic} conditions which are at the heart of this work.

We find that uniaxial pressure along the $c$ axis significantly
reduces the transition pressures in both systems.  {\Ca} shows for
both pressure conditions an orthorhombic to collapsed tetragonal
transition, though the transition is less abrupt when uniaxial
pressure is applied. For {\Ba} we observe two phase transitions from
orthorhombic to collapsed tetragonal through an intermediate
nonmagnetic tetragonal phase.  An analysis of the electronic
bandstructure features near the critical pressures reveals the origin
of the sensitivity of the systems to pressure conditions.

\begin{figure*}[tb]
  \begin{center}
    \includegraphics[width=0.92\textwidth]{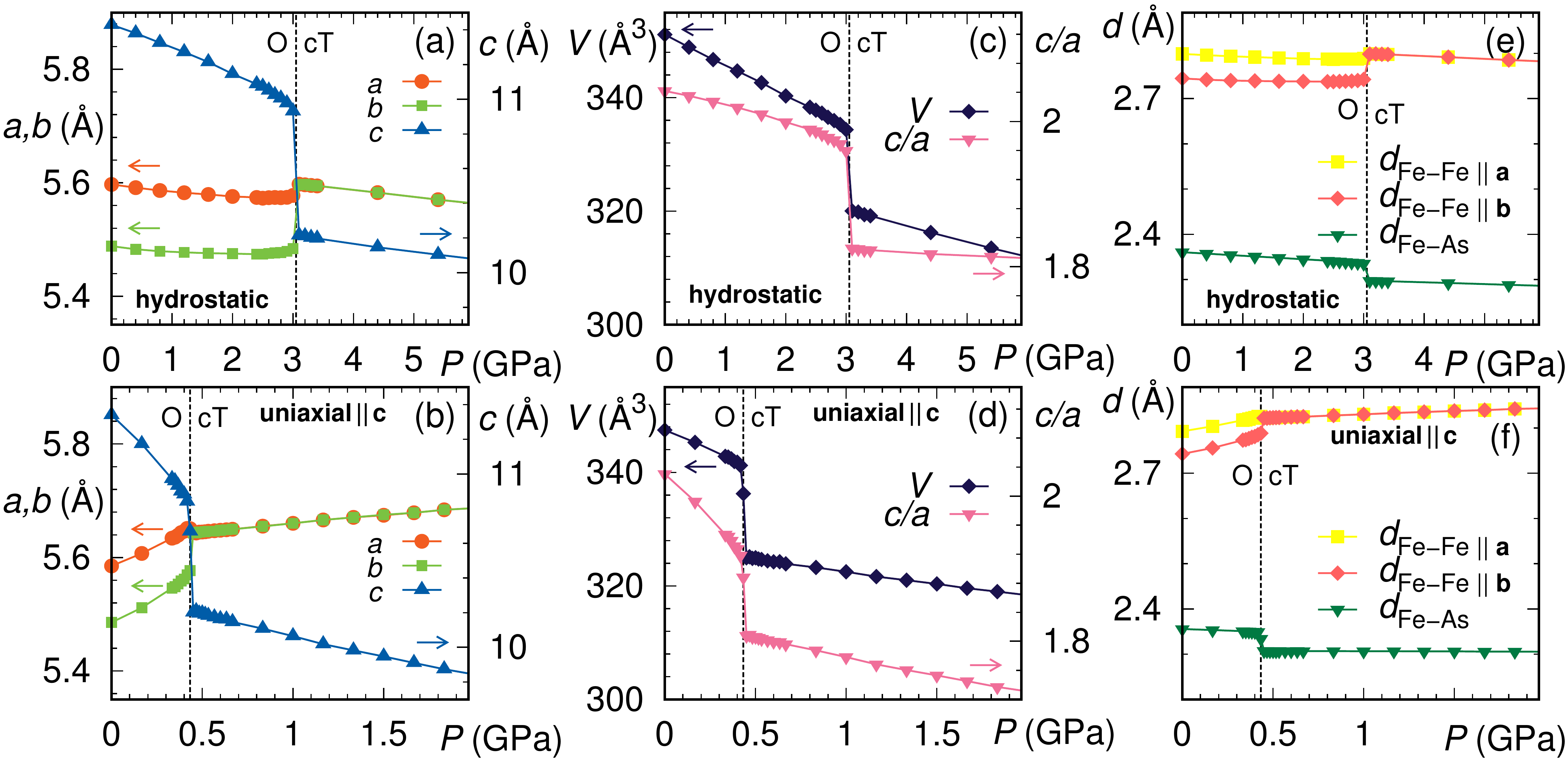}
    \caption{Structure of {\Ca} under hydrostatic and uniaxial pressure.
      Lattice parameters, a) and b), volume and axis ratio, c) and d),
      c) and d), selected bond lengths e) and f).}
      \label{ca_struct}
  \end{center}
\end{figure*}

Calculations were performed using the Vienna ab initio simulations
package (VASP)~\cite{Kresse1993} with the projector augmented wave
(PAW) basis~\cite{Bloechl1994} in the generalized gradient
approximation (GGA). Structural relaxations under hydrostatic pressure
were carried out with the conjugate gradient (CG) method as
implemented in the VASP package.
 The energy cutoff was set to 300~eV and a
  Monkhorst-Pack uniform grid of $(6\times 6\times 6)$ points was used
  for the integration of the irreducible Brillouin zone. For
  relaxations with the CG algorithm two cycles were performed in order
  to minimize the error caused by the Pulay stress. Note that the
  reported bond compressions of up to 7\% at 50~GPa don't affect the
  precision of the PAW basis.
 In order to
perform relaxations under uniaxial pressure we modified the fast
inertial relaxation engine~\cite{Bitzek2006} (FIRE) algorithm to be
able to handle full structural relaxations with an arbitrary stress
tensor.

In Fig.~\ref{ca_struct} we show the evolution of lattice parameters,
volume and Fe-Fe, Fe-As distances under hydrostatic and ($c$-axis)
uniaxial pressure for {\Ca}.  We find a first order phase
transition from a magnetic (stripe order) orthorhombic phase to a
nonmagnetic collapsed tetragonal phase at $P_c=3.05$~GPa ($P_c=0.48$~GPa)
under hydrostatic (uniaxial) pressure and at zero temperature we don't
observe any intermediate tetragonal phase under uniaxial
stress~\cite{Prokes2010}.

In hydrostatic conditions, $a$ and $b$ expand as a consequence of the
Pauli principle~\cite{Zhang2009} with $b$ abruptly increasing in
value, while $c$ shows a significant collapse of 6.5{\%}
(Fig.~\ref{ca_struct}~(a)) and the unit cell volume shows a sharp drop
of about 4.3{\%} (Fig.~\ref{ca_struct}~(c)).  The value of
$c/a_t=2.58$ with $a_t=a/\sqrt{2}$, indicates the onset of a collapsed
tetragonal phase.  Our results are in good qualitative agreement with
experimental~\cite{Kreyssig2008} observations, except for the
overestimation of the critical pressure ($P_c^{\rm exp}=0.3$~GPa) also
found in previous theoretical studies~\cite{Zhang2009,Colonna2011}.
Following the changes of the lattice parameters at $P_c$, the inplane
Fe-Fe distances show a sharp increase at $P_c$ while the out-of-plane
Fe-As distance decreases (Fig.~\ref{ca_struct}~(e)).  Using the
generalized Birch-Murnaghan $p-V$ equation of state~\cite{Sata2002} we
obtained a bulk modulus $B=70\pm3$~GPa at ambient pressure, while at
$P_c$ the bulk modulus jumps from $56\pm3$ to $105\pm2$~GPa.
In order to obtain these estimates we performed a series
of fits for every phase separately 
considering every pressure point of our data as a reference
pressure. In this way we obtain the bulk modulus as a function of
pressure.



In contrast to the hydrostatic case, when uniaxial pressure is applied
(Fig.~\ref{ca_struct} (b)) the $a$ and $b$ lattice parameters expand
significantly while $c$ is compressed up to $P_c=0.48$~GPa where a
drop for $c$ is observed while $a$ and $b$ continue to expand
monotonously. The volume reduces by 3.4{\%} and the ratio $c/a_t=2.56$
at $P_c$ (Fig.~\ref{ca_struct}~(c)) denotes the entrance to a
collapsed tetragonal phase, where magnetism is suppressed
completely. The phase transition shifts to smaller $P_c$ compared to
hydrostatic pressure, which is in very good agreement with experiments
under nonhydrostatic pressure conditions~\cite{Prokes2010}.
Nevertheless, the authors of Ref.~\onlinecite{Prokes2010} find for
pressures above 0.06~GPa a stabilization of the high-temperature
tetragonal structure down to temperatures below the superconducting
transition. This phase is not seen in our calculations which may be
related to the fact that at very low temperatures the tetragonal phase
may be disappearing again (Fig. 1 of Ref.~\onlinecite{Prokes2010}).

\begin{figure*}[htb]
  \begin{center}
    \includegraphics[width=0.8\textwidth]{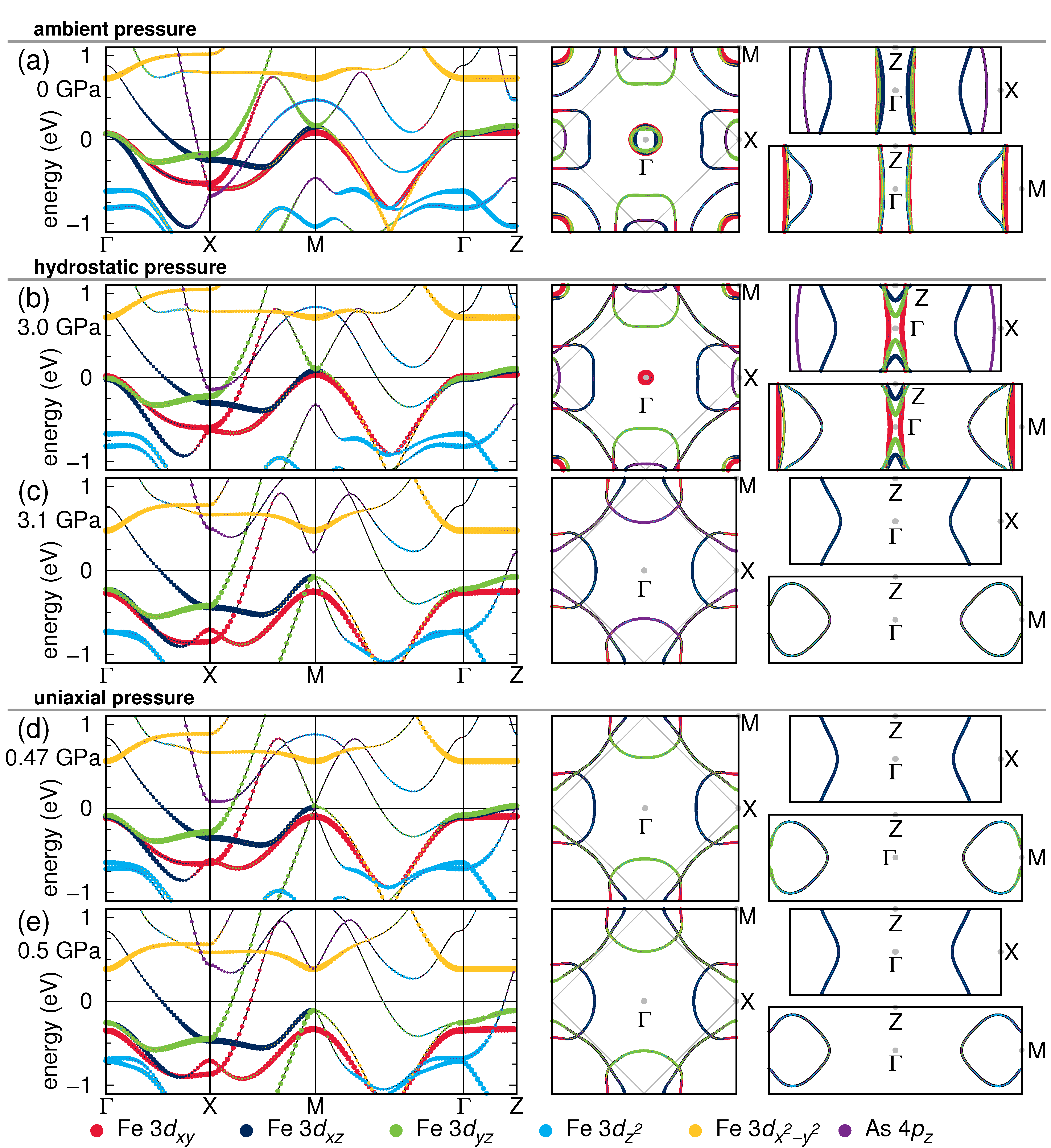}
    \caption{Bandstructure and $k_z=0$, $k_y=0$ and $k_x+k_y=0$ Fermi
      surface cuts of {\Ca}~\cite{FPLO}. For the orbital character,
      $x$ and $y$ point along the nearest neighbour Fe-Fe
      connections.}
      \label{ca_bs}
  \end{center}
\end{figure*}

In order to understand the differences in behavior observed between
the hydrostatic and uniaxial pressures, we show in Fig.~\ref{ca_bs}
the orbital weighted bandstructure and $k_z=0$, $k_y=0$ and
$k_x+k_y=0$ Fermi surface cuts of {\Ca} under hydrostatic
(Fig.~\ref{ca_bs} (b)-(c)) and uniaxial pressure (Fig.~\ref{ca_bs}
(d)-(e)) at pressures below and above the phase
transition. Bandstructures and Fermi surfaces were calculated using the full-potential local orbital (FPLO) basis~\cite{FPLO}.
The bandstructure and Fermi surface cuts at ambient
pressure are also shown for comparison (Fig.~\ref{ca_bs} (a)). We use
the orthorhombic space group $F\,mmm$ for all band structure plots in
order to facilitate comparison. The behavior of the electronic
structure in the vicinity of the Fermi energy is crucial for
understanding the transition. Right below $P_c$ both pressure
conditions show a high density of Fe $d_{xz}$, $d_{yz}$ and
$d_{x^2-y^2}$ states at $E_F$ (see in Fig.~\ref{ca_bs} (b) and (d) the
$\Gamma$-Z path and near M) which are pushed away above $P_c$
(Fig.~\ref{ca_bs} (c) and (e)) and the hole pockets at $\Gamma$
disappear, suppressing possible nesting conditions.  The compression
along $c$ enforces the interlayer As~$p_z$-As~$p_z$
bonding~\cite{Johrendt1997} which can be related to the proximity of
the As~$p_z$ band to $E_F$ near $P_c$.  Uniaxial stress is for this
process more effective than hydrostatic pressure since a similar
electronic behavior is reached at much smaller pressures as observed
in Fig.~\ref{ca_bs} (d)-(e). In agreement with
Ref.~\onlinecite{Yildirim2009} the collapsed tetragonal phase sets in
as soon as the Fe magnetic moment goes to zero. Also, the shape of the
Fermi surface derived from Fig.~\ref{ca_bs} in the collapsed
tetragonal phase agrees well with the de Haas van Alphen measurements
performed for CaFe$_2$P$_2$ ($c/a_t$ = 2.59) where a highly dispersive
topology in the $c$ axis as well as the absence of the hole pocket at
the $\Gamma$ point has been reported~\cite{Coldea2009} (compare
Fig. ~\ref{ca_bs} (c) and ~\ref{ca_bs} (e) with Figs. 2 and 3 of
Ref.~\cite{Coldea2009}). The isoelectronic substitution of As by P in
{\Ca} corresponds to application of chemical pressure and shows
similar features to the collapsed tetragonal phase of {\Ca} obtained
after application of (hydrostatic) pressure. The similarity of
chemical pressure and applied pressure has already been discussed in
Refs.~\onlinecite{Coldea2009} and \onlinecite{Kimber2009}. In fact,
comparison of our obtained $c/a_t= 2.58$ ratio and As position
$z_{As}=0.1358$ (hydrostatic) in the collapsed tetragonal phase of
{\Ca} with the measured $c/a_t=2.59$ and P position $z_{P}=0.1357$ of
CaFe$_2$P$_2$ shows the high resemblance between both crystal
structures.

\begin{figure*}[bt]
  \begin{center}
    \includegraphics[width=0.92\textwidth]{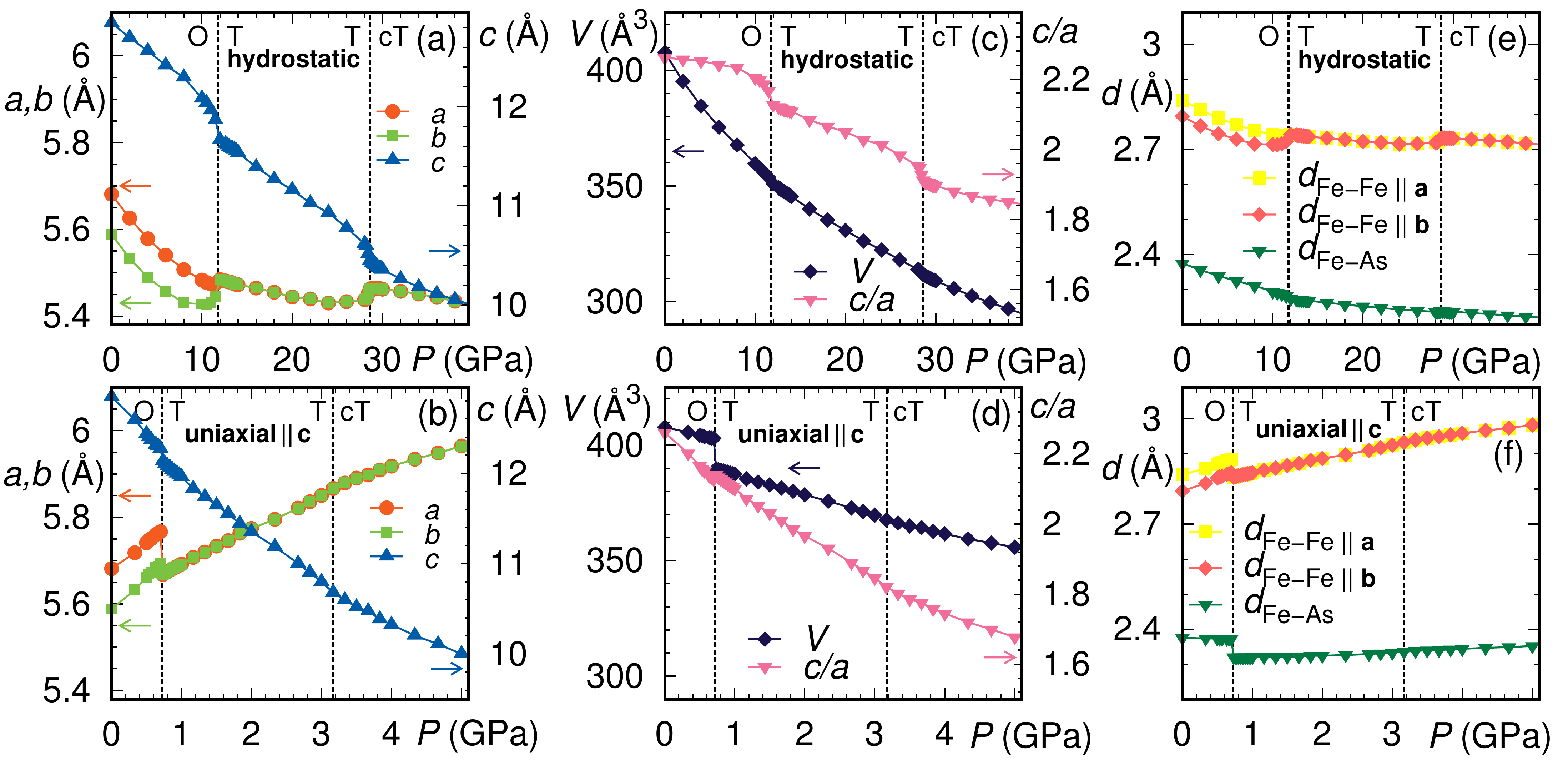}
    \caption{Structure of {\Ba} under hydrostatic and uniaxial
      pressure.  Lattice parameters, a) and b), volume and axis ratio,
      c) and d), selected bond lengths e) and f).}
      \label{ba_struct}
  \end{center}
\end{figure*}

\begin{figure}[htb]
  \begin{center}
    \includegraphics[width=0.42\textwidth]{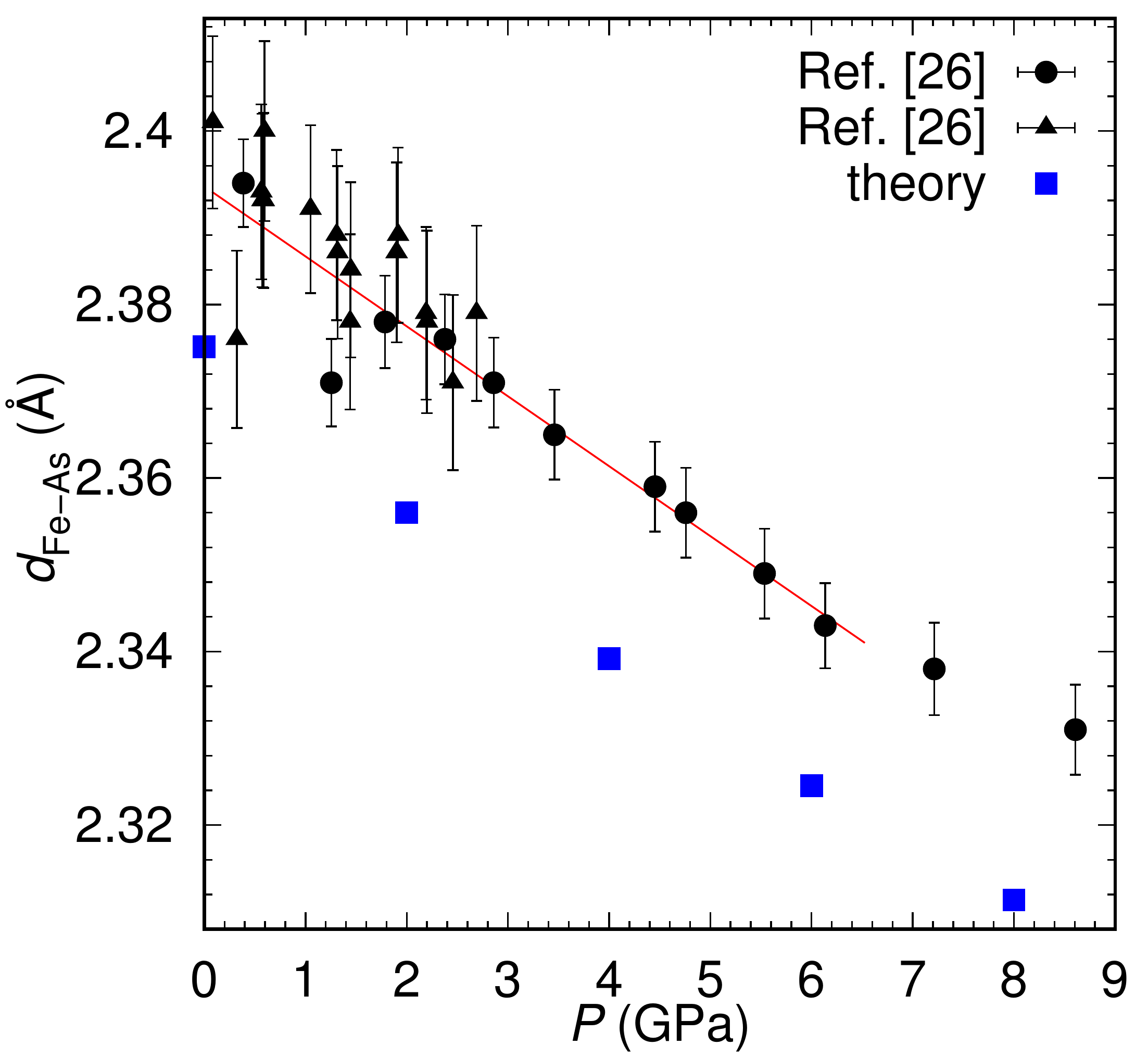}
    \caption{Comparison of $T=0$ calculated (squares) and $T=298$~K
      measured (triangles, circles)~\cite{Granado2011} Fe-As bond
      distances in {\Ba} as a function of pressure.}
      \label{Fe_Asbonds}
  \end{center}
\end{figure}

We now proceed with the analysis of {\Ba}.  In Fig.~\ref{ba_struct} we
present the changes in lattice parameters, volume and atomic distances
under hydrostatic and uniaxial pressures for {\Ba}. Similar to {\Ca},
the critical pressures under uniaxial stress are reduced with respect
to hydrostatic conditions. This observation was also reported by
recent constant volume density functional theory calculations on {\Ba}
under nonhydrostatic pressure~\cite{Colonna2011_2}.  {\Ba}, contrary
to {\Ca}, shows two phase transitions.  At $P_{c_1}=11.75$~GPa
($P_{c_1}=0.72$~GPa) we find a phase transition from an
antiferromagnetic orthorhombic to a nonmagnetic tetragonal phase under
hydrostatic (uniaxial) conditions.  A second smooth phase transition
to a collapsed tetragonal phase is obtained for $P_{c_2}=28.6$~GPa
($P_{c_2}=3.17$~GPa) (Fig.~\ref{ba_struct} (c)-(d))~\cite{Uhoya2010}.
High-pressure neutron diffraction experiments~\cite{Kimber2009} as
well as previous theoretical calculations under hydrostatic pressure
conditions also find a phase transition to an intermediate tetragonal
phase~\cite{Zhang2009,Colonna2011} but recent syncrotron X-ray
diffraction experiments under nonhydrostatic conditions see no
signature of an intermediate tetragonal phase at low temperatures.
Nevertheless, an anomaly in the As-Fe-As bond angles at $P\sim
10$~GPa~\cite{Mittal2011} as well as a loss of magnetic
moment~\cite{Duncan2010} have been reported.  This could be related to
the phase transition that we find at $P_{c_1}=11.75$~GPa where
magnetism is suppressed.  At higher pressures the agreement of the
onset of the collapsed tetragonal phase with the X-ray diffraction
data~\cite{Mittal2011} is very good. Clearly the phase transitions in
{\Ba} are less abrupt than in {\Ca}.
The ambient pressure bulk modulus is estimated at $67\pm4$~GPa, in
good agreement with experimentally reported values \cite{Mittal2011}
of $82.9\pm1.4$ and $65.7\pm0.8$~GPa at 33~K and 300~K,
respectively. At $P_{c_1}$, the bulk modulus abruptly increases from
$98\pm4$ to $128\pm3$~GPa and at $P_{c_2}$ it jumps from $150\pm3$ to
$173\pm2$~GPa. This is in very good agreement with the experimental
estimate of $B=153\pm3$GPa for the collapsed tetragonal
phase~\cite{Mittal2011}.  We also analyzed the Fe-As bond
compressibility for $P= 9$~GPa (hydrostatic) and found
$\kappa=3.5\times 10^{-3}$~GPa$^{-1}$ which is in excellent agreement
with $\kappa=3.3\times 10^{-3}$~GPa$^{-1}$ obtained in an extended
X-ray absorption fine structure (EXAFS) experiment~\cite{Granado2011}.
In Fig.~\ref{Fe_Asbonds} we show the comparison of the measured
pressure dependence of the Fe-As bond distances~\cite{Granado2011}
with our results. Due to different temperatures (experiment is performed
at room temperature, theory at $T=0$) our distances are shorter by about
0.02~{\AA} (0.8\%), but the overall agreement is good.

\begin{figure*}[htb]
  \begin{center}
    \includegraphics[width=0.8\textwidth]{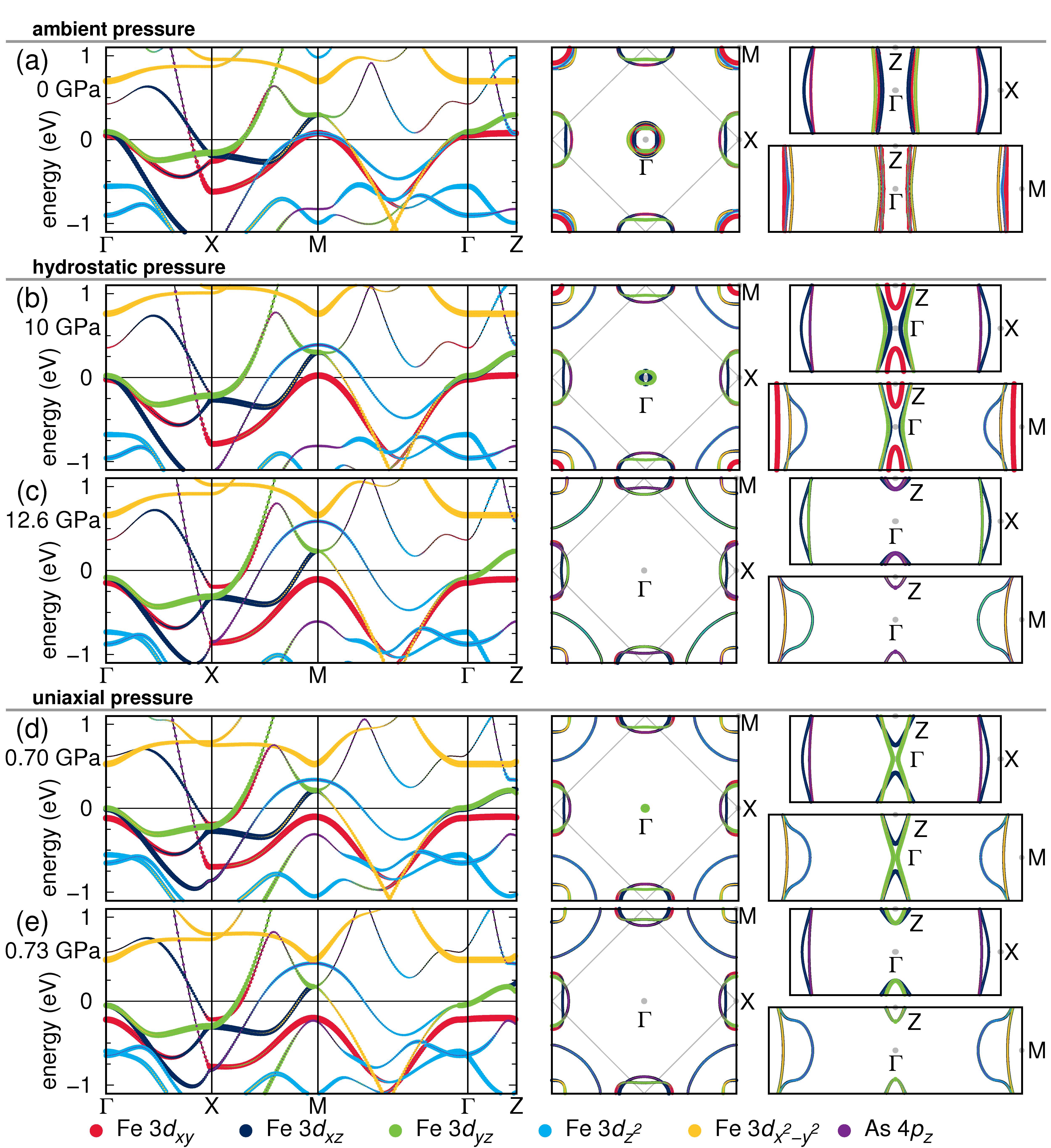}
    \caption{Bandstructure and $k_z=0$, $k_y=0$ and $k_x+k_y=0$ Fermi
      surface cuts of {\Ba}~\cite{FPLO}.}
      \label{ba_bs}
  \end{center}
\end{figure*}

In Fig.~\ref{ba_bs} we present the orbital weighted bandstructure
and $k_z=0$, $k_y=0$ and $k_x+k_y=0$ Fermi surface cuts of
{\Ba} under hydrostatic (Fig.~\ref{ba_bs} (b)-(c)) and uniaxial
pressure conditions (Fig.~\ref{ba_bs} (d)-(e)) at two pressures
below and above the orthorhombic to tetragonal phase transition at
$P_{c_1}=11.75$~GPa ($P_{c_1}=0.72$~GPa).  Similar to {\Ca} we observe
below $P_{c_1}$ a high density of Fe $d_{xz}$, $d_{yz}$ and
$d_{x^2-y^2}$ states at $E_F$ which is pushed down (less drastically
than in {\Ca}) for pressures above $P_{c_1}$.
The hole pockets disappear at the $\Gamma$ point and the Fe magnetic
moment goes to zero.  Here the As $p_z$ band seems to be little
affected at the critical pressure.  In contrast, at $P_{c_2}=28.6$~GPa
($P_{c_2}=3.17$~GPa) (bandstructure not shown) the As~$p_z$ band is
pushed towards the Fermi level indicating a strong As~$p_z$-As~$p_z$
bonding while entering the collapsed tetragonal phase.  These results
show that under perfect hydrostatic or perfect uniaxial pressure
conditions neither the intermediate tetragonal phase nor the collapsed
tetragonal phase fulfill Fermi surface 
nesting conditions.  In fact, we find that the structural
parameters measured in Ref.~\onlinecite{Kimber2009} are similar to our
calculated parameters far below $P_{c_1}$ in the orthorhombic phase
(except for the orthorhombic distorsion), where well defined hole
pockets are found at the $\Gamma$ point.

In summary, we have presented finite pressure density functional
theory calculations which allow the investigation of {\it
  nonhydrostatic} pressure conditions. Our finite pressure relaxed
structures show good agreement with the available experimental
data (volume, Birch-Murnaghan values and compressibilities) though our
magnetic moments in the orthorhombic phase are larger than the
observed experimental values.  We expect that this overestimation
affects mostly the values of $P_c$.  Comparison of our calculated
Fe-As bond distances at different pressures with measured
distances~\cite{Granado2011} shows good agreement. Also,
available de Haas van Alphen measurements performed for
CaFe$_2$P$_2$~\cite{Coldea2009} agree well with our predicted Fermi
surface shapes for {\Ca} in the collapsed tetragonal phase.  This
overall agreement with experimental observations demonstrates that the
presented constant pressure calculations provide a reliable
theoretical prediction of structures under nonhydrostatic pressure
conditions, allowing for arbitrary stress tensors in future
studies. Such calculations can complement experiments and help
identify the precise degree of hydrostaticity.  We find that uniaxial
stress along the $c$ axis considerably reduces the critical pressures
for {\Ca} and \Ba.  This behavior can be understood by the fact that
the phase transitions are strongly dictated by the electronic
properties in the vicinity of the Fermi energy, as shown by our
electronic structure analysis. While {\Ca} undergoes a magnetic
orthorhombic to a non-magnetic collapsed tetragonal phase for both
pressure conditions and no indication of an intermediate tetragonal
phase is observed under uniaxial stress, {\Ba} shows two phase
transitions from a magnetic orthorhombic to a collapsed tetragonal
phase through an intermediate nonmagnetic tetragonal phase for both
pressure conditions.  All nonmagnetic phases show a disappearance of
the hole pockets at the $\Gamma$ point suppressing possible Cooper
pair scattering channels between electron and hole pockets.  Such
scattering channels have been discussed to be important for the
superconductivity in BaFe$_2$As$_2$~\cite{Graser10}.  More experiments
need to be done in order to understand the origin of the
superconducting phase observed in these materials under various
pressure conditions.

{\bf Acknowledgments.-} We would like to thank A. I. Coldea for
useful discussions, the Deutsche
Forschungsgemeinschaft for financial support through grant SPP 1458,
 the Helmholtz Association for support through HA216/EMMI and
the centre for scientific computing (CSC, LOEWE-CSC) in Frankfurt
for computing facilities.

\end{document}